# Noise Resilient Exceptional-Point Voltmeters based on Neuromorphic functionalities

Arunn Suntharalingam[1], Lucas Fernández-Alcázar[2], Rodion Kononchuck[1], Tsampikos Kottos[1]

[1]Wave Transport in Complex Systems Lab, Department of Physics, Wesleyan University, Middletown CT, USA

[2]Instituto de Modelado e Innovación Tecnológica (CONICET-UNNE) and Facultad de Ciencias Exactas, Naturales y Agrimensura, Universidad Nacional del Nordeste, W3404AAS Corrientes, Argentina

**Abstract:** Exceptional point degeneracies (EPD) of linear non-Hermitian systems have been recently utilized for hypersensitive sensing. This proposal exploits the sublinear response that the degenerate frequencies experience once the system is externally perturbed. The enhanced sensitivity, however, might be offset by excess (fundamental and/or technical) noise. Here, we developed a self-oscillating nonlinear platform that supports transitions between two distinct neuromorphic functionalities – one having a spatially symmetric steady-state, and the other with an asymmetric steady-state – and displays nonlinear EPDs (NLEPDs) that can be employed for noise-resilient sensing. The experimental setup incorporates a nonlinear electronic dimer with voltage-sensitive coupling and demonstrates two-orders signal-to-noise enhancement of voltage variation measurements near NLEPDs. Our results resolve a long-standing debate on the efficacy of EPD-sensing in active systems above self-oscillating threshold.

The underlying mathematical structures of non-Hermitian wave systems [1][2][3][4] have inspired the last few years new technologies [5][6][7][8]. Many of these are reliant on the existence of exceptional point degeneracies (EPDs) [8]. These are non-Hermitian degeneracies where a set of N eigenvalues and their corresponding eigenvectors coalesce [1][2]. In the proximity of an $N$−th order EPD (EPD-N), the eigenvalue detuning $\Delta f \equiv |f − f_{EPD}|$, due to a small external perturbation ε, follows a sublinear response (SLR) $\Delta f \sim \sqrt[N]{\varepsilon} \gg \varepsilon$ that can be utilized for enhanced sensing [9][10][11][12][13][14][15].

A principal requirement for efficient EPD sensing is the increase of the resolution limit via the narrowing of the resonance linewidth. This can be achieved by a judicious design of cavity amplification mechanisms. The downside of this strategy is that it introduces additional noise that, in some EPD lasing platforms, might offset the enhanced signal sensitivity [13][16][17][18][19]. Furthermore, nonlinear effects might become important, requesting the development of theoretical tools that treat them on equal footing with the sensitivity enhancement near EPDs. Most current studies, however, rely on linear mathematical constructs, such as the Petermann factor [20][21][22], which describes the linewidth enhancement near EPDs due to the bi-orthogonal nature of the eigenmodes of the underlying linear non-Hermitian systems[16]. Obviously, this approach is not suitable when the response of a system is influenced by nonlinearities. An example case is a laser at an EPD. Fortunately, an appropriate language exists from the area of dynamical systems and bifurcation theory [23][24][25][26][27], which can be adopted for the analysis of nonlinear EPDs (NLEPDs). Examples of systems that are amenable to such analysis are shown in Fig. 1A. In fact, some recent theoretical studies have utilized this approach to address issues like the formation of NLEPDs and the emulation of neuronal dynamic functions using parity-time (PT) symmetric systems that involve gain and loss nonlinear channels [26][27]. Among these neuromorphic functionalities, are oscillation quenching mechanisms whose characteristics are determined by the underlying *dynamical symmetries* of the system. Two distinct oscillation quenching mechanisms have been singled-out [28][29][30]: The first is the so-called oscillation death (OD) [28], which is associated with a spatially asymmetric steady-state and leads to an explicit parity-time (PT) symmetry violation, see Fig. 1B.I. The other mechanism, known as amplitude death (AD) [29], results to a spatially uniform steady-state which is protected by a dynamical (self-induced) PT-symmetry, see Fig. 1B.II. OD constitutes a well-known phenomenon in neurons, known as the "winner-take-all" situation, whereas AD mainly serves to suppress neuronal oscillations [30]. The transition from AD to OD might be characterized by the formation of a NLEPD (see Fig. 1B right and Fig. 1C), and it is a generic phenomenon found in a plethora of physical systems (e.g. see Fig. 1A). Can these NLEPDs be used for sensing and what is the signal-to-noise enhancement (SNE) factor in their proximity? A definite answer to this question requires not only a theoretical modeling [24][25][31][32][33][34][35], but most importantly the establishment of controllable experimental platforms where the predictions of the theory will be scrutinized and guide the theoretical language as it is developing.

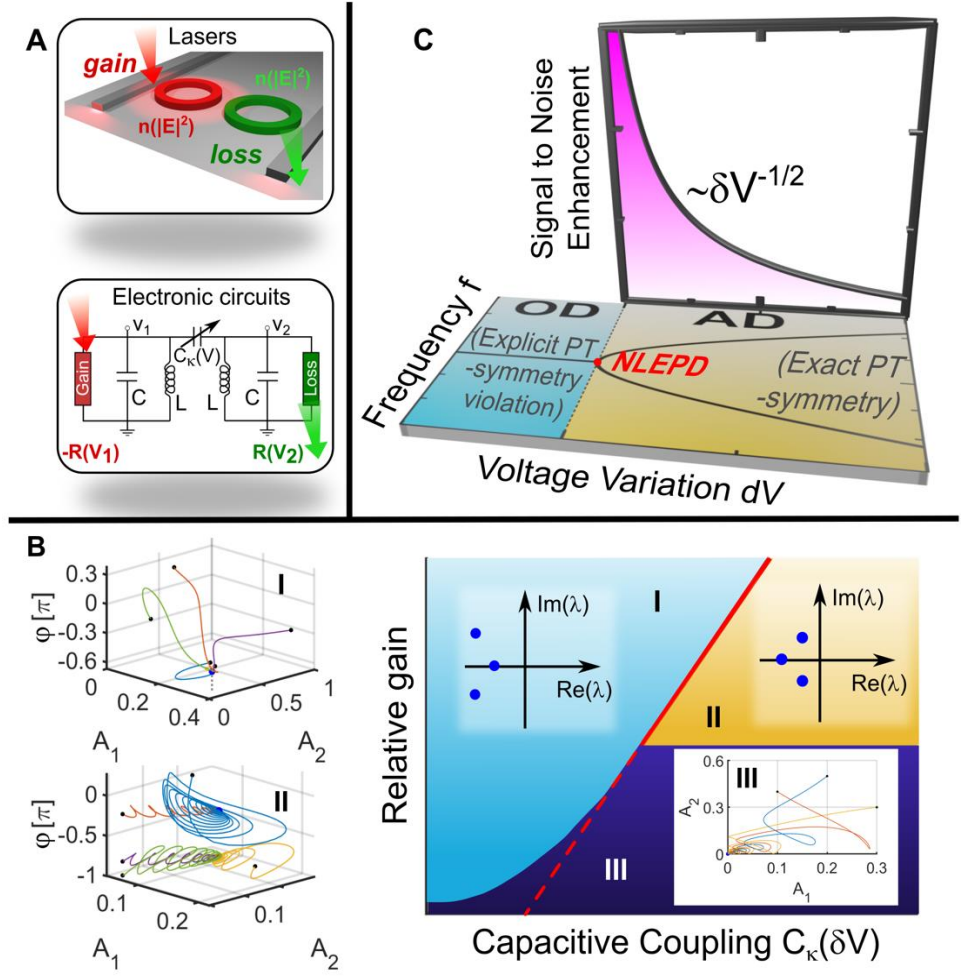

Figure 1: **EPDs at the transition between OD and AD.** (a) Physical systems that demonstrate OD and AD oscillation quenching. (b) (Right) The parameter space of the circuit of the subfigure of Fig. 1.A, is partitioned in three distinct domains that host stable NS (fixed points of the dynamical equations associated with Jacobian eigenvalues $\{\lambda_n; n = 1,2,3\}$ with $\mathcal{R}(\lambda_n) < 0$) with distinct dynamical symmetries. Transition from one domain to another is dictated by the relative gain $\gamma_1^{(0)}/\gamma_2^{(0)}$ and voltage variation $\delta V$ that controls the capacitive coupling between the two nonlinear RLC tanks. The three domains are: (I) the oscillation death (OD) domain, (II) the amplitude death (AD) domain, and (III) the trivial steady-state solution domain. The steady-state field amplitude of each resonator differs from one another in domain (I) while it is the same in domain (II). Typical examples of phase-space trajectories in cases (I) and (II) are shown on the left subfigures. The black dots indicate initial conditions while the blue dots the steady-state (fixed point). The red solid line indicates NLEPDs associated with the coalescence of two stable NS. The red dashed line indicates NLEPDs associated with the coalescence of two non-trivial unstable NS (see Supplementary Information) (c) (Horizontal plane) Parametric evolution of the nonlinear eigenfrequencies $f$ versus $\delta V$ for a fixed value of the relative gain corresponding to a transition from OD to AD via a nonlinear EPD (NLEPD), see red dot. (Vertical plane) The signal-to-noise enhancement factor $SNE = \chi/\alpha_{VRW}$ (where $\chi \equiv \partial(\Delta f_+/f_0)/\partial(\delta V)$ is the sensitivity and $\alpha_{VRW}$ is the noise-equivalent voltage variations) diverges as $1/\sqrt{\delta V}$ in the proximity of NLEPD.

Here, we address the viability of NLEPD sensing protocols using two nonlinear RLC tanks whose capacitive coupling is used as a sensing platform for voltage variations. The RLC circuits have anharmonic parts consisting of a complementary amplifier (gain) and a dissipative conductor (loss), see Fig. 1A and Methods. The nonlinear supermodes (NS) are the fixed points of the dynamical equations, and their properties arise from the underlying dynamical symmetries and their stability. We focus on stable NS that are experimentally accessible and are identified from the (negative) real part of the eigenvalues $\{\lambda_n\}$ of the Jacobian matrix, which describes the linearized dynamics around each of these fixed points [23]. Their properties lead to the partition of the parameter space in three distinct domains (see domains I, II, III in Fig. 1B right): The last domain III involves trivial NS with zero amplitude at each RLC tank (see inset of right subfigure of Fig. 1B) and is therefore irrelevant to our investigations. The other two, are separated by a NLEPD (see red line in right subfigure of Fig. 1B and red point in Fig. 1C) and contain one (two) non-trivial stable hyperbolic fixed points in the OD (AD) phase, see Fig. 1C (and Fig. 1B.I and Fig. 1B.II respectively). The AD stable NSs coalesce at the NLEPD-point at a voltage variation $\delta V = 0$, where $\delta V$ controls the capacitive coupling between the two resonators. The detuned eigenfrequencies follow a characteristic SLR $\Delta f_\pm \equiv f_\pm - f_{NLEPD} \propto \pm\sqrt{\delta V}$, leading to two-orders enhancement of sensitivity to small voltage variations and a similar SNE near the NLEPD, see Fig. 1C. Our results challenge the validity of linear concepts (e.g. Petermann factor) for the noise analysis near NLEPDs and confirm beyond doubt that self-oscillating systems above threshold have an enhanced signal-to-noise sensing performance in the proximity of the NLEPDs.

**Experimental Platform:** The sensor consists of a pair of nonlinear RLC resonators [36][37][38][39] (see Fig. 1A) with natural frequency $f_0 = \frac{1}{2\pi}\frac{1}{\sqrt{LC}} \approx 338 KHz$ and impedance (at resonance) $Z_0 = \sqrt{L/C} \approx 424\ \Omega$. One of the resonators (gain—indicated with red in Fig. 1A) incorporates a nonlinear amplifier $-R_1(V_1)$ characterized by an I-V curve $I_1(V_1) = -V_1/R_1^{(0)} + bV_1^3$ while the other one (loss—indicated with green in Fig. 1A) incorporates a nonlinear loss $R_2(V_2)$ with an I-V curve $I_2(V_2) = V_2/R_2^{(0)} + bV_2^3$ ($b \approx 7 \cdot 10^{-4} A/V^3$, and $V_{1(2)}$ are the voltages at the nodes 1(2)). The two resonators are coupled together via a Capacitance Voltage Controlled (CVC) capacitor $C_\kappa(V) = \kappa \cdot C$ where $\kappa$ is a dimensionless parameter representing the strength of the coupling. The linear conductances $\frac{1}{R_1^{(0)}} > \frac{1}{R_2^{(0)}}$ were tuned such that the system undergoes a transition from AD to OD as the voltage at the coupling capacitor varies (see Figs. 1B,1C and Methods). A transmission line (TL) with impedance $z_0 = 50\ \Omega$ is weakly coupled to each resonator via capacitors $C_e \ll C$. The TLs were used to collect the signal generated by the circuit and direct it to a VNA for further processing (see Methods). The NLEPD (occurring at $\delta V$=0) can be experimentally identified as the point for which the voltage amplitudes $V_1/V_2$ of each RLC resonator deviates from unity (AD domain) acquiring larger values (OD domain), see Fig. 2A.

**Theoretical Analysis of Oscillation Quenching and Characterization of Nonlinear Supermodes:** The voltage dynamics $V_n$ at each RLC tank ($n = 1,2$), is described via a temporal coupled mode theory (TCMT) (see Supplementary Information)

$$i \frac{d}{d\tau} \begin{pmatrix} a_1 \\ a_2 \end{pmatrix} = \begin{pmatrix} \nu_\kappa + i\gamma_1 & \frac{\kappa}{2} \\ \frac{\kappa}{2} & \nu_\kappa - i\gamma_2 \end{pmatrix} \begin{pmatrix} a_1 \\ a_2 \end{pmatrix}; \quad (1)$$

where $a_n \equiv \sqrt{\frac{3}{8} b Z_0} \left( V_n + i \frac{1}{2\pi} \frac{\dot{V}_n}{f_0} \right)$, $\gamma_n = \gamma_n^{(0)} + (-1)^n (|a_n|^2 + \eta)$, $\nu_\kappa = 1 - \frac{\kappa}{2} - \sqrt{\frac{\eta}{2} \frac{Z_0}{Z_0}}$ and $\tau \equiv 2\pi f_0 t$ is the rescaled time. The parameters $\gamma_n^{(0)}$ are the linear gain ($n = 1$) and loss ($n = 2$) coefficients associated with the gain and loss RLC resonator respectively, while $\eta \equiv \frac{1}{2} \frac{Z_0}{Z_0} \left( \frac{C_e}{C} \right)^2$ models the coupling of the circuit to the TLs. The global frequency shift $f_\kappa = f_0 \cdot \nu_k$ is associated with the renormalization of the natural frequency $f_0$ of the RLC resonators due to the capacitive coupling between them ($\kappa$ −term) and the coupling with the TLs ($\eta$-term). Its $\kappa$-dependence could be (in principle) avoided if we choose another type of coupling (e.g. inductive coupling). Below, we analyze the steady-state properties of Eq. (1) in terms of the coupling parameter $\kappa = \kappa(\delta V)$, which is used as a sensing platform of voltage variations $\delta V$ (see Methods).

  The nonlinearities in our system have been chosen carefully to prevent the system from evolving towards undesirable unbounded states where $A_1$, and/or $A_2 \to \infty$. This can be easily realized from Eq. (1) by recognizing that whenever the field intensity of the gain resonator exceeds a critical value $|a_1|^2 > \gamma_1^{(0)} - \eta$ the gain coefficient $\gamma_1$ becomes negative, thus turning the gain RLC tank into a lossy one. The NS of Eq. (1) may be expressed in the polar representation $a_n = A_n e^{i\varphi_n} e^{-if\tau/f_0}$ and are identified as the fixed points of the dynamical system Eq. (1) whose evolution is defined in a three-dimensional phase space $(A_1, A_2, \varphi \equiv \varphi_2 - \varphi_1)$. We classify these fixed points according to their underlying (dynamical) symmetry, and their stability. The latter is determined by the eigenvalues $\{\lambda_1, \lambda_2, \lambda_3\}$ of the $3 \times 3$ Jacobian matrix $J$ when it is evaluated at the fixed point (see Supplementary Information) [23]. When $Re(\lambda_n) \neq 0$ ($\forall n = 1,2,3$), the fixed point is a hyperbolic equilibrium and there is a homeomorphism that maps the phase portrait in its proximity onto solutions of its linearized system described by $J$ [40]. When all $Re(\lambda_n) < 0$, the fixed point is stable while it is unstable if at least one $Re(\lambda_n) > 0$. Hyperbolic equilibria are robust to small variations which do not change (qualitatively) the phase portrait. The opposite scenario of non-hyperbolic equilibria is associated with cases where one of the eigenvalues of the Jacobian matrix is zero or has zero real part. These are structurally unstable cases, and one can numerically test the nature of the stability of these fixed points by direct dynamical simulations with Eq. (1).

We have found that one fixed-point of Eq. (1) is a trivial state $(A_1, A_2) = (0,0)$. The analysis of the Jacobian eigenvalues (see Supplementary Information) indicates that it is stable in the parametric domain III (see Fig. 1B) while it is unstable in the other two domains. Below, we analyze the stable non-trivial fixed points occurring in the domains I and II in the Fig. 1B. In these cases, the real-valued amplitudes $A_n > 0$ take the form (see Supplementary Information):

$$A_n = \rho^{n-1} \cdot \sqrt{\gamma_1^{(0)} - \eta - \frac{\kappa\rho}{2}}; \quad \text{for } \kappa \leq \gamma_1^{(0)} + \gamma_2^{(0)} \text{ (domain I)},$$
$$A_1^{(\pm)} = A_2^{(\pm)} = \sqrt{\frac{\gamma_1^{(0)} - \gamma_2^{(0)} - 2\eta}{2}}, \text{ for } \kappa \geq \gamma_1^{(0)} + \gamma_2^{(0)} \text{ (domain II)}, \quad (2)$$

where the real-valued variable $\rho \equiv \frac{A_2}{A_1} > 0$ is a solution of the quartic algebraic equation

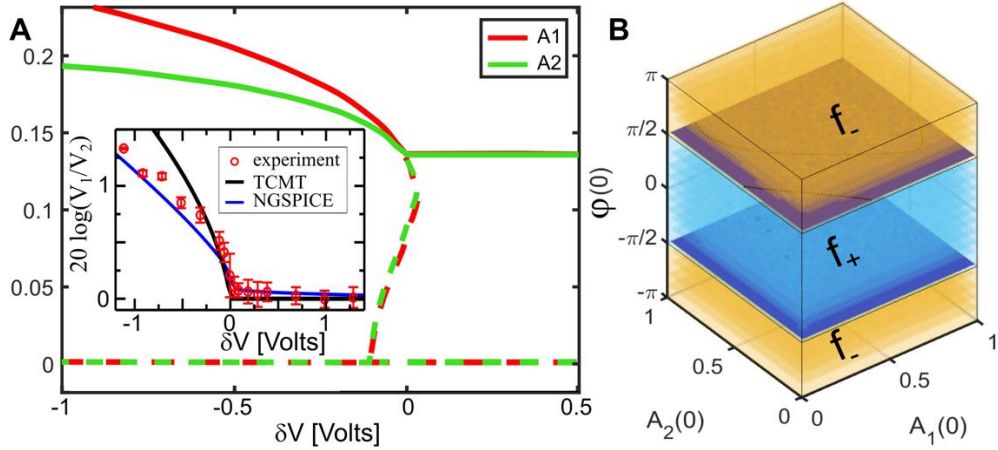

Figure 2: **Stability Analysis and Basins of Attraction:** (A) Field amplitude $A_1(A_2)$ of the NS versus voltage variations $\delta V$. The solid lines indicate stable fixed points (see Eq. (2)) while the dashed lines indicate unstable solutions evaluated numerically using Eq. (1) together with the eigenvalue analysis of the Jacobian matrix (see Supplementary Information). The fixed relative gain $\gamma_1^{(0)}/\gamma_2^{(0)} = 1.46$ is chosen in a way that the system undergoes a transition from OD to AD as the voltage variation $\delta V$ increases. The NLEPD occurs at $\delta V = 0$. (Inset) Measured average logarithmic voltage ratio (red symbols) $V_1/V_2$ of the NS. Each point represents an average of five independent measurements and the error bars have been extracted from error analysis of the $V_1$ and $V_2$ measurements. The black line is the prediction of TCMT. The blue line indicates the results from NGSPICE. (B) Phase-space analysis and basins of attraction for the stable fixed points associated with the upper ($f_+$) (blue highlighted domain) and lower ($f_-$) (yellow highlighted domain) branches of the NS in the AD domain. The voltage variation is $\delta V \approx 2\ mV$ corresponding to a circuit configuration in the proximity of the NLEPD.

$1 - 2\rho\left(\frac{\gamma_2^{(0)}+\eta}{\kappa}\right) - 2\rho^3\left(\frac{\gamma_1^{(0)}-\eta}{\kappa}\right) + \rho^4 = 0$. The physical requirement $A_{1,2}^{(\pm)} \in \Re$ ($A_n \in \Re$) leads to the condition $\gamma_1^{(0)} - \gamma_2^{(0)} - 2\eta \geq 0$ ($\gamma_1^{(0)} - \eta - \frac{\kappa\rho}{2} \geq 0$) which determines the boundary between the domains II and III (domains I and III), see Fig. 1B. Finally, the relation $\kappa_{NLEPD} = \gamma_1^{(0)} + \gamma_2^{(0)}$ defines the transition between domains I and II and which is characterized by the formation of a NLEPD associated with the coalescence of two stable NS (see solid red line in Fig. 1B). This is further confirmed by evaluating the nonlinear eigenfrequencies $f_\pm(\kappa)$ associated with the solutions of Eq. (2). From the TCMT we have that (see Supplementary Information):

$$f_\pm = \begin{cases} f_\kappa & \text{for } k \leq \kappa_{NLEPD} \text{ (domain I)} \\ f_\kappa \pm \frac{f_0}{2}\sqrt{\kappa^2 - \kappa_{NLEPD}^2}, & \text{for } k \geq \kappa_{NLEPD} \text{ (domain II)} \end{cases}, \quad (3)$$

where the square-root dependence of the eigenfrequencies from the coupling detuning $\kappa$ reflects the presence of the NLEPD. At the range of voltage variations that have been used in our experiment ($-1.1 \leq \delta V \leq 2[V]$ with resolution of 1[mV]), the coupling is $\kappa(\delta V) \approx \kappa_{NLEPD} - 0.0234[\frac{1}{V}] \cdot \delta V[V]$ with $\kappa_{NLEPD} \equiv \gamma_1^{(0)} + \gamma_2^{(0)} \approx 0.30$ (where $\gamma_1^{(0)} \approx 0.18$; $\gamma_2^{(0)} \approx 0.12$).

The two fixed points in domain II ($k \geq \kappa_{NLEPD}$), are associated with the AD phase where the field amplitudes at each resonator are the same and the two stable NS differ only by the relative phase $\varphi_\pm$ (see Supplementary Information). In this parameter range, the system respects an exact parity-time symmetry i.e., both the system and the corresponding NS are invariant under a joint parity (i.e., space inversion $1 \leftrightarrow 2$) and time-reversal (i.e., complex conjugation) symmetry. In domain I ($k \leq \kappa_{NLEPD}$), instead, there is only one non-trivial stable NS. This domain is associated with the OD phase where the field amplitudes at each RLC resonator differ from one-another. A detailed fixed point numerical analysis using a MATLAB fsolve routine (see Fig. 2A) confirms the above theoretical predictions and provides a more general information about other (unstable) fixed points as well. In the inset of Fig. 2A, we also report some represented values of the measured voltage ratios $V_1/V_2$ versus the voltage variation $\delta V$. These results compare nicely with the TCMT predictions (black line) indicating that the NLEPD occurs at the transition between AD and OD phases. The deviations for large negative $\delta V$ are attributed to the limitations of the TCMT and/or small detunings of various components of our circuit from its ideal (TCMT) parameters. At the same figure, we also report the numerical simulations from NGSPICE (blue solid line) nicely capturing the experimental results.

The existence of a stable NS does not guarantee the evolution of the system to this specific state. Instead, the system may evolve either to a stable trivial state, or to another stable NS in case of bistabilities (AD domain). The former scenario is easily excluded by an appropriate choice of the relative gain parameter (see Fig. 1B). The latter scenario can be controlled by realizing that the final state depends strongly on the initial conditions $\{A_1(0), A_2(0), \varphi(0)\}$. The phase-space volume that contains initial conditions which converge to a specific stable fixed point constitute its basin of attraction, and its size provides a measure of how attractive this fixed point is. Detailed dynamical simulations using Eq. (1), for various $\kappa$ −values and with a fine mesh of initial conditions $\{A_1, A_2, \varphi\}$, allowed us to identify the basins of attraction of the two fixed points in the AD domain. We find that initial excitations with a relative phase $|\varphi| > \frac{\pi}{2}$ end up at the AD fixed point associated with the $f_-$ mode while an initial preparation with a phase $|\varphi| < \frac{\pi}{2}$ leads to a $f_+$ supermode. In Fig. 2B we show the basins of attraction for the $f_+(f_-)$ fixed points which are indicated with blue (yellow) color for the example case of $\delta\kappa \equiv \kappa(\delta V) - \kappa_{NLEPD} \approx 5 \cdot 10^{-5}$ (corresponding to $\delta V \approx 2\ m$Volts). In fact, further analysis using both NGSPICE and TCMT indicated that an appropriate detuning between the resonant frequencies of the two RLC resonators destroys the bistability by favoring only the upper fixed point $f_+$ when the system is very close to the NLEPD. Away from the NLEPD the bistable nature at the AD domain persists. Either way, the square-root scaling of the NS frequency $f_+$ from $f_{NLEPD}$ (see Eq. (3)) is unaffected.

**Sensing Protocol:** In Fig. 3A we report a density-plot of the voltage power spectrum $|V_1(\omega)|^2$ for various voltage variations $\delta V$ by performing a Fourier transform of the temporal field, evaluated via time-domain simulations of the TCMT Eq (1). To achieve the asymptotic states associated with the $f_+(f_-)$ supermodes in the AD domain, we have prepared the initial excitation at relative phase $|\varphi| < \frac{\pi}{2}$ ($|\varphi| > \frac{\pi}{2}$) as discussed above. The numerical data agree nicely with the theoretical predictions of Eq. (3). In Fig. 3B we show a density plot of the measured power spectrum of the emitted signal together with the TCMT predictions of Eq. (3) for the $f_+$ frequency (red dashed line). The absence of $f_-$ from the measured power spectrum, is associated with the fact that the experimental initial preparation favors a field excitation with a small relative phase $|\varphi| < \frac{\pi}{2}$. At the same figure we also report the $f_+$ frequency (versus $\delta V$) that has been extracted by a Fourier transform of the voltage $V_1(t)$ using NGSPICE simulations (black dotted line).

The sublinear detuning is better appreciated by reporting $\Delta f_+ \equiv f_+ - f_{NLEPD}$ vs. $\delta V$ in a double-logarithmic plot. The experimental data (cyan cicrles) nicely match the results from the NGSPICE (dashed black line) showing the predicted behavior $\Delta f_+ \propto \sqrt{\delta V}$ from TCMT, see Fig. 3C. This sublinear response offers an opportunity to develop an enhanced sensing protocol for detecting small voltage variations $\delta V$ using as a sensing platform the coupling capacitor $C_\kappa(\delta V)$. At the same time, the square-root SLR extends the dynamical range (DR) of the sensing measurements up to relatively large values of $\delta V$. The DR is the other important metric that characterizes the efficiency of a sensor, and it is defined as the ratio between the maximum and the minimum $\delta V$ variation that the sensor can measure. Furthermore, the presence of gain elements guarantees the narrowing of the emission peaks and promotes an enhanced resolution. To further quantify the efficiency of our sensing protocol, we have introduced the sensitivity $\chi \equiv \partial(\Delta f/f_0)/\partial(\delta V)$. In Fig. 3D we report the measured sensitivity (violet circles) together with the NGSPICE results (black dashed line). We find that $\chi \sim 1/\sqrt{\delta V}$ in the proximity of the

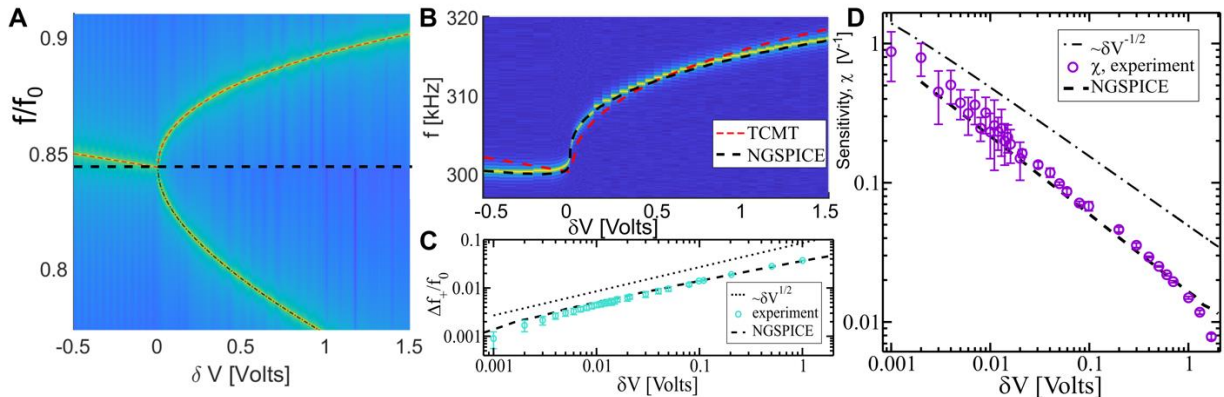

Figure 3: **Experimentally measured sublinear frequency detuning and sensitivity to applied voltage variations:** (a) Density plot of the normalized emitted spectrum evaluated from dynamical simulations of the TCMT model versus voltage variations of the coupling capacitor. The nonlinear frequencies for $\delta V > 0$ have been obtained using different initial conditions which belong to the basin of attraction of the corresponding stable fixed point. The dashed black line indicates the frequency domain for which each initial condition has been used. (b) Measured emitted spectrum as a function of voltage variations. The red dashed line in both (a-b) is the TCMT prediction Eq. (3) of the non-linear frequencies. (c) The measured relative frequency detunings (circles) for the stable fixed point associated with the upper branch versus the applied voltage variations. The black dashed line is drawn to guide the eye and has a slope 1/2 characteristic of a NLEPD of order N=2. (d) The sensitivity of the active nonlinear circuit demonstrating two orders enhancement in the proximity of the NLEPD as opposed to a system configuration away from the NLEPD. The black dashed line in subfigures B,C,D indicates the numerical results using NGSPICE. Error bars in subfigures C and D indicate $\pm 1$ standard deviation obtained from ten independent measurements.

NLEPD.

**Noise Analysis:** The sublinear frequency detuning Eq. (3) guarantees an enhanced transduction function from the voltage variation to the sensitivity $\chi$. It does not, however, addresses another important characteristic of high-performance sensors that is related to the precision of the measurements. The latter is identified with the smallest measurable variation in the input signal that can be identified by the sensor due to noise at the output signal.

To better understand the effects of noise in the measurement process we have analyzed the Allan deviation $\sigma_{\widetilde{\Delta f}_+}(\tau)$ of the normalized frequency detunings $\widetilde{\Delta f}_+ \equiv \Delta f_+/f_0$ as a function of the sampling time $\tau$. The measured Allan deviation is reported in Fig. 4A for representative $\delta V$ values – both well within the NLEPD-enhanced sensitivity regime and away from it. We observe that as $\delta V$ decreases, and the system approaches towards the NLEPD, the noise increases. To better appreciate the effects of noise in the measured voltage variation, we report in Fig. 4B the normalized Allan deviation $\sigma_\alpha(\tau) = \sigma_{\widetilde{\Delta f}_+}(\tau)/\chi$ [Volt]. Our measurements show that the noise grows slower than the signal enhancement as we are approaching the NLEPD. We conclude, therefore, that the proposed NLEPD protocol can provide an enhanced SNR in its proximity. Based on the behavior of Allan deviation, we can distinguish different regimes depending on the duration of the sampling time $\tau$. Each regime is influenced by a different type of noise sources. At the limiting case of long sampling times, the Allan deviation behaves as $\sigma_\alpha^{DRR} = \alpha_{DRR}\tau$. This is typical of a drift rate ramp (DRR) noise associated with the presence of systematic (deterministic) errors. For intermediate sampling times, the Allan deviation reaches a saturation value $\sigma_\alpha^{BI}(\tau) = \alpha_{BI}\tau^0$ which is indicative of the bias instability (BI) noise. The value of $\alpha_{BI}$ sets the smallest possible reading of our sensor. Its origin is traced to the random flickering of electronics or other components of the system. Finally, the short-time behavior of Allan deviation exhibits voltage (variations) random walk (VRW) noise which decreases with the sampling time $\tau$ as $\sigma_\alpha^{VRW} = \alpha_{VRW} \cdot \tau^{-1/2}$. The origin of the VRW is traced to noise sources such as the thermal (Johnson-Nyquist) noise from the circuit elements (like the resistors and the amplifier) and from the attached TL, the readout noise, or other noise sources associated with fluctuations of the voltage applied to the coupling capacitor or capacitance fluctuations due to thermal variations. Each of them is described by a separate noise coefficient $\alpha_{cir}, \alpha_{TL}, \alpha_{det}, \alpha_{add}$ respectively, and contributes to noise equivalent voltage variation as $\alpha_{VRW} = \sqrt{\alpha_{JN}^2 + \alpha_{det}^2 + \alpha_{add}^2}$, where $\alpha_{JN}^2 = \alpha_{cir}^2 + \alpha_{TL}^2$. Since $\alpha_{det}$ can be actively minimized, the thermal noise $\alpha_{JN}$, together with $\alpha_{add}$ represent the best obtainable limit for $\alpha_{VRW}$.

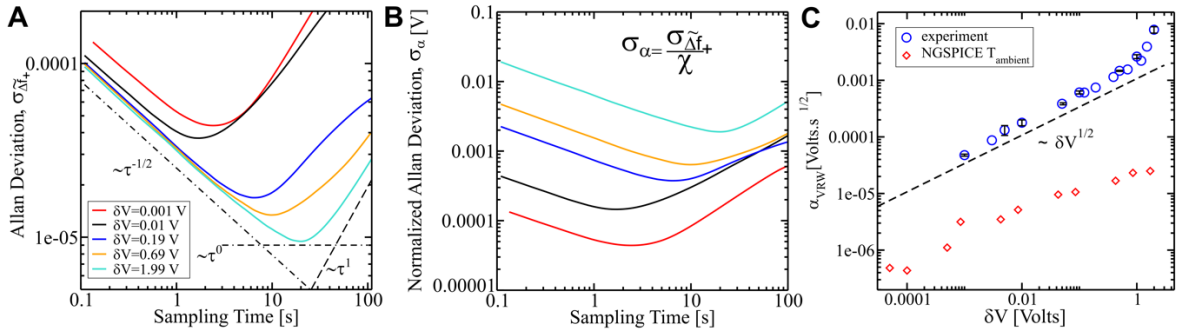

Figure 4: **Noise Analysis at various voltage variations:** (a) The Allan deviation $\sigma_{\widetilde{\Delta f}_+}(\tau)$ of the circuit readout versus the sampling time $\tau$ is measured at various voltage variations $\delta V$ of the coupling capacitor both in the proximity (small $\delta V$-values) and away (large $\delta V$-values) from the NLEPD. (b) The rescaled (with respect to the sensitivity $\chi$) Allan deviation $\sigma_\alpha(\tau) = \sigma_{\widetilde{\Delta f}_+}(\tau)/\chi$ decreases as we are approaching the NLEPD indicating that the sensitivity enhancement offsets the noise enhancement. (c) The measured (blue circles) voltage random walk coefficient $\alpha_{VRW}$ versus the voltage variations for all $\delta V$ voltage variations that we have used. For some representative $\delta V$-values we also indicate error bars representing $\pm 1$ standard deviation evaluated over four different measurements. The red diamonds are the results of NGSPICE simulations where we have considered thermal (Johnson-Nyquist) noise at the resistors, amplifiers and at the TLs described by an ambient temperature $T = 300\ K$.

In Fig. 4C we provide a panorama of $\alpha_{VRW}$ for all $\delta V$-values that we have used in our measurements (blue circles). From the measurements we conclude that $\alpha_{VRW} \approx 0.002 \, \delta V^{0.5} \, [Volt][Hz]^{0.5}$ indicating a robustness to the VRW noise, which is attributed to the stable hyperbolic nature of the $f_+$ supermode. Specifically, the phase-space around these hyperbolic fixed points is *structurally* stable [40], and the associated basin of attraction is relatively broad – even for voltage variations that are extremely close to the NLEPD (see Fig. 2B). Therefore, the VRW noise is not able to "push out" of the basin of attraction the phase-space trajectories. At the same figure we also show the noise coefficient $\alpha_{JN}$ from NGSPICE (red diamonds) where we have incorporated thermal noise at the resistors, amplifier and at the TLs described by an ambient temperature $T = 300 \, K$. This is more than one-order of magnitude smaller than our experimental measurements for $\alpha_{VRW}$. We conclude that the detection noise $\alpha_{det}$ combined with $\alpha_{add}$ overwhelms the thermal noise $\alpha_{JN}$. Eventually, the absolute bound of $\alpha_{VRW}$ will be determined by the multiplicative noise due to coupling fluctuations originating from voltage uncertainties and capacitance fluctuations due to temperature variations.

**Conclusion:** We have demonstrated a SNE sensing of an NLEPD-based voltmeter. The proposed sensing protocol is based on a square-root frequency detuning from the NLEPD induced by a small voltage variation modifies the coupling between two nonlinear RLC tanks. The NLEPD occurs at the transition between two types of oscillation quenching regimes i.e., OD and AD domains in the parameter space and is a consequence of the nonlinear gain/loss channels assigned to each RLC tank. The nonlinearity leads to a structurally stable phase space in the proximity of the stable fixed points associated with the degenerate NS, while the corresponding basins of attraction are relatively broad, even for voltage variations that are extremely close to the NLEPD. These characteristics, shield the sensing signal from noise and lead to a two-orders enhancement of signal-to-noise ratio in the proximity of the NLEPD. Our results establish the EPD-sensing from self-oscillating systems above threshold as an efficient platform with dramatically improved SNE factor in the proximity of the NLEPD. Our scheme can guide the design of novel neuromorphic hypersensitive sensors with enhanced dynamical range that can be utilized in electroencephalography, electrocardiography and neuroprosthetics.

**Acknowledgements:** We acknowledge useful discussions with Prof. U. Kuhl on improvements of noise analysis, Prof. F. Ellis on circuit design and Mr. W. Tuxbury for assisting with the experimental platform.

**Methods:**

**Circuit Design and Fabrication:** The circuit schematic used for this experiment can be seen in Extended Data Fig. 1 below. It features two *RLC* resonators which are coupled to each other via voltage-controlled capacitors. The main elements that make up each *RLC* resonator are a resistor, $R_i$, an inductor, $L_i$; and a pair of in-parallel grounded capacitors, $C_v$ and $C_i$, where $i = 1,2$, denotes the gain and loss *RLC* units respectively. The inductors, $L_i = 200\,\mu H$, used in both the gain and loss resonators are API Delevan 807-1537-90HTR. The total capacitance in each resonator is made up by a combination of a tunable capacitor, $C_v$, connected in parallel with a fixed capacitor, $C_i$. The tunable capacitor, $C_v$, is a Murata 81-LXRW19V201-058 with a capacitance range of $100 - 200\,pF$. A voltage of $0.5V$ was applied to the tunable capacitor in each *RLC* resonator and controlled by a EG&G Instruments 7265 DSP lock-in amplifier, that was connected via a Bayonet Neill–Concelman (BNC) port to a resistor, $R_v$, model Yageo 603-RC0402FR-074K99L, with a resistance of $R_v = 4.99\,k\Omega$. This resistor was connected to a grounded fixed capacitor, $C_{v1}$, a Murata 81-GCM32EL8EH106KA7L, with a capacitance of $C_{v1} = 10\,\mu F$. The fixed capacitors in each resonator unit, $C_i$, is a Kemet 80-C0603C911F5G with a capacitance of $910\,pF$. This gives a total capacitance in each *RLC* resonator of $200\,pF + 910\,pF = 1{,}110\,pF$.

Each *RLC* resonator has resistive elements that collectively provide gain and loss. The former is provided by an operational amplifier (op-amp) model Analog Devices 584-ADA4862-3YRZ-R7. The power supply for the op-amp was connected by a standard 3 pin connector, with one going to ground, and the other two being connected to $V_+ = 6V$ and $V_- = -6V$ respectively. To produce gain, the op-amp has a pair of internal resistances, $R_{G1}$ and $R_{G2}$, of $550\Omega$ each. $R_{G1}$ is connected in between the output of the op-amp and the inverting input of the op-amp. $R_{G2}$ is connected on one end to the inverting input of the op-amp and grounded on the other end. The mechanically tunable variable resistor in the gain *RLC* tank, $R_1$, is a Vishay 71-PHPA1206E2001BST1 component which has a resistance of $R_1 = 2000\Omega$. $R_1$ is connected on one end to the operational amplifier's non-inverting input. The other end of $R_1$ is connected to a capacitor, $C_{e1}$, model Kemet 80-C0805C100FDTACTU where $C_{e1} = 10\,pF$, capacitively couples each *RLC* resonator to transmission lines. A mechanically tunable variable resistor, $R_{T1}$, is connected in parallel to $R_1$ and the op-amp, is a Bourns 652-3269W-1-102GLF with $R_{Ti} = 720 \pm 200\Omega$. The set value of $R_{T1} = 720\Omega$, and it is connected to a pair of grounded back-to-back of diodes, $D_i$, Onsemi 512-1N914BWS – which represent the nonlinear elements of this *RLC* dimer. The fixed resistor in the loss resonator, $R_2$, is a Vishay 71-PCNM2512E2501BST5 component with a resistance of $R_2 = 2500\Omega$. $R_2$ is connected in parallel to $R_{T2} = 750\Omega$. On the other end, $R_{T2}$ is connected to a pair of back-to-back grounded diodes – the same model of diodes as in the gain resonator was used.

The coupling between the two resonators was achieved using two parallel variable capacitors, $C_{vc}$. The component used in $C_v$, a Murata 81-LXRW19V201-058 is the same model as the variable capacitors in the resonator units, $C_{vc}$. As in the $RLC$ tanks, $C_{vc}$ is connected to a grounded $C_{v1}$ which in turn is connected to $R_v$. The same lock-in amplifier is used to control the tuning voltage via a BNC port. One of the two capacitors is held fixed at $200 pF$ with a constant applied voltage of $0.5V$, whereas the other is tuned in voltage range between $0.4 - 3.5V$.

**Voltage and frequency detuning measurements**

The emitted signal from the electronic circuit was collected for different applied voltage variations of the Capacitance Voltage Control ($C_{VC}$) capacitor. These voltage variations where electronically controlled via an EG&G Instruments 7265 DSP lock-in amplifier. The imposed voltage variations were in the range between 0.4-3.5V associated with $-1.1 \leq \delta V \leq 2[V]$ with resolution of up to 1[mV]. At each specific voltage variation, the emitted spectrum was collected using a network analyzer Keysight E5080A. The individual frequency sweeps contain 4001 points in a range of 295-320 kHz. A single measurement was obtained from the collected spectrum with an intermediate frequency bandwidth (IFBW) of 100Hz giving a sampling time of 40.01s. The peak frequencies of the spectrum $f_+$ were then identified from the resulted spectrum, which allows the calculation of the frequency detuning $\Delta f_+$.

**Allan deviation measurement**

The Allan deviation $\sigma_{\widetilde{\Delta f}_+}(\tau)$ of the frequency associated with the voltage power spectrum peak of the emitted-signal is defined as

$$\sigma_{\widetilde{\Delta f}_+}(\tau) = \sqrt{\frac{1}{2(M-1)} \sum_{n=1}^{M-1} \left( \langle \widetilde{\Delta f}_+^{(n+1)} \rangle - \langle \widetilde{\Delta f}_+^{(n)} \rangle \right)^2}, (4)$$

where $\tau$ is the sampling time, $M$ is the total number of frequency measurements and $\langle \widetilde{\Delta f}_+^{(n)} \rangle$ indicates the average rescaled frequency detuning during the sampling time interval $[n\tau, (n+1)\tau]$. For the extraction of Allan deviation, the rescaled emitted peak detunings $\widetilde{\Delta f}_+ \equiv \Delta f_+/f_0$ were sampled with an IFBW of 10 kHz for 101 points in a frequency range of 7kHz centered around the expected $\Delta f_+$ for the associated $\delta V$. 20,000 consecutive spectral measurements were performed over a period of approximately 2700s for each $\delta V$. This results in a sampling time of 0.1337s.

**NGSPICE simulations**

We use NGSPICE, an open-source software for electronic circuits, to simulate the dynamical behavior of our experimental platform. We consider two $RLC$ tanks coupled by a capacitor using the same characteristics as the experimental platform, where, unless specified otherwise, we use the same parameters of the electronic components as described in 'Circuit Design and Fabrication' section of the Methods. The op-amp in the gain resonator is represented by a high impedance Norton amplifier by designating its constituent components – a transconductance that quantifies gain, a capacitor and diode clippers. Nonlinearity in the gain and loss resonators is modeled with back-to-back diodes via 1N914 diodes - the same type as used in the experimental platform- using the appropriate parameters that describe its behavior. To compensate for the detuning due to the capacitor in the op-amp, the capacitor in the gain resonator, $C_v + C_1$, was detuned slightly to $0.9955(C_v + C_1)$, to fit the experimental data obtained. In conjunction with that $L_i$ (i = 1,2) was set at $0.965L_i$, in the simulations. Using this setup, we evaluate the signal generated by the circuit for a total time of $t = \frac{6000}{f_0} \approx 0.017s$ with the time steps of $dt = \frac{1}{(80f_0)} \approx 37ns$. In our analysis of the power spectrum, we dropped the first 0.0017s, which correspond to the short time transient, to consider only the steady-state behavior.

To add noise, we modeled Johnson-Nyquist noise – electronic noise due to thermal fluctuations. This was added to our simulations by adding random voltage sources at every resistor in our circuit (including the TLs). The amount of noise added at each resistor using the TRNOISE function of NGSPICE at these resistors are dictated by the root mean square of the voltage due to Johnson-Nyquist noise. This is given by $v_{rms} = \sqrt{4k_BTRB}$ where $k_B$ is Boltzmann's constant; $R$ is the value of the respective resistor at which the random voltage source was added; $B$ is the bandwidth of the noise for which the natural frequency of the resonator $\approx 338\ kHz$ was assumed; and $T$ is the temperature for which the ambient value of $300\ K$ was chosen.

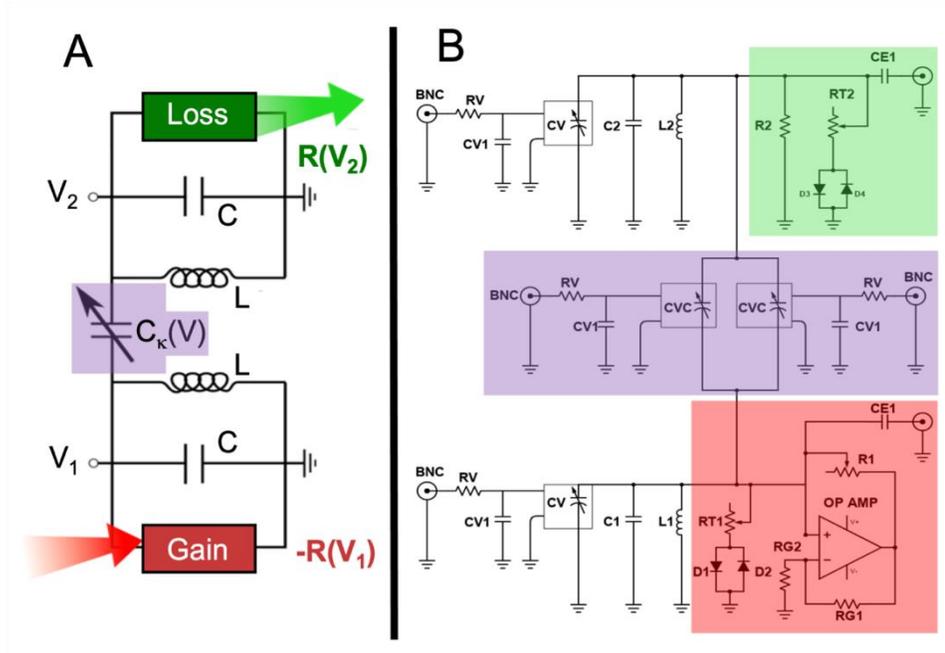

**Extended Data Fig. 1**: **Schematic of the circuit**: (A) The conceptual circuit (B) The actual circuit diagram. Various parts are highlighted with the same color as in (A).

# Noise Resilient Exceptional-Point Voltmeters based on Neuromorphic functionalities

Arunn Suntharalingam[1], Lucas Fernández-Alcázar[2], Rodion Kononchuck[1], Tsampikos Kottos[1]

[1]Wave Transport in Complex Systems Lab, Department of Physics, Wesleyan University, Middletown CT, USA
[2]Instituto de Modelado e Innovación Tecnológica (CONICET-UNNE) and Facultad de Ciencias Exactas, Naturales y Agrimensura, Universidad Nacional del Nordeste, W3404AAS Corrientes, Argentina

**Supplementary Information**

**Coupled Mode Theory modeling of the electronic circuit**

The nonlinear electronic circuit of Fig. 1A of the main text can be described by the temporal coupled mode theory (TCMT), see Eq. (1) of the main text. Apart from the simplicity and the physical intuition that a TCMT modeling can offer, it also allows us to extrapolate our findings to a broader class of systems described by such TCMT. We start the derivation of Eq. (1) by expressing the energy stored in the LC resonator as $|\psi|^2$ where the complex modal amplitude of a single LC resonator $\psi(t)$ is

$$\psi(t) = \sqrt{\frac{C}{2}} \left( V(t) - \frac{i}{\omega_0} \dot{V}(t) \right). \qquad (S1)$$

Above $V(t)$ is the node voltage and $\omega_0 = 2\pi f_0 = 1/\sqrt{LC}$ the resonant frequency of the resonator in the absence of dissipation (amplification). Using is representation, we rewrite the circuit equation of the LC resonator, i.e., $\frac{d^2}{dt^2} V + \omega_0^2 V = 0$, as a set of two (uncoupled) first order differential equations i.e., $\frac{d\psi}{dt} = i\omega_0 \psi$ and its complex conjugate. The latter relation emphasizes the time dependence of the modal amplitude $\psi \sim e^{i\omega_0 t}$ and $\psi^* \sim e^{-i\omega_0 t}$.

Next, we incorporate in our circuit, an energy dissipation (enhancement) associated with a nonlinear damping (antidamping) channel. We consider nonlinear components that are characterized by the I-V curve $I(V) = \alpha V/R^{(0)} + b V^3$ where $\alpha = 1$ ($\alpha = -1$) represents loss (gain) and $b \approx 7 \cdot 10^{-4} [\frac{A}{V^3}]$. The corresponding Kirchoff's equation for the individual RLC resonator collapse to the following expression

$$\ddot{V}(t) + \alpha \Gamma^{(0)} \dot{V}(t) + \beta V^2(t) \dot{V}(t) + \omega_0^2 V(t) = 0, \qquad (S2)$$

where $\Gamma^{(0)} = (R^{(0)} C)^{-1}$ is the relaxation rate of the individual (linear) LRC oscillator, and $\beta = 3b/C$. Equation (S2) can be written in terms of $\psi$ and its conjugate $\psi^*$ as

$$2\dot{\psi}^* + 2i\omega_0 \psi^* - (\alpha \Gamma^{(0)} + \frac{\beta}{2C}|\psi|^2)(\psi - \psi^*) - \frac{\beta}{2C}[\psi^3 - (\psi^*)^3] = 0. \qquad (S3)$$

By invoking a rotating wave approximation and eliminating the fast-oscillating terms, the above equation can be further simplified as

$$i\dot{\psi}^* \approx \omega_0 \psi^* - i\frac{\alpha}{2}\Gamma^{(0)}\psi^* - i\chi|\psi^*|^2\psi^*, \quad (S4)$$

where $\chi = \beta/(4C) = 3b/(4C^2)$.

We proceed by developing a TCMT associated to the two coupled RLC tanks. From Kirchoff's laws we obtain the equations for the voltage $V_n$ in resonator $n = 1,2$

$$\begin{aligned}(1+\kappa)\ddot{V}_1 - \kappa\ddot{V}_2 - \Gamma_1^{(0)}\dot{V}_1 + \beta V_1^2\dot{V}_1 + \omega_0^2 V_1 &= 0\\ (1+\kappa)\ddot{V}_2 - \kappa\ddot{V}_1 + \Gamma_2^{(0)}\dot{V}_2 + \beta V_2^2\dot{V}_2 + \omega_0^2 V_2 &= 0\end{aligned}, \quad (S5)$$

where $\Gamma_n^{(0)} = \left(R_n^{(0)}C\right)^{-1}$ is the relaxation rate of each resonator (below we will assume the high-Q limit, i.e., $\frac{\Gamma_n^{(0)}}{\omega_0} \ll 1$), $C_\kappa = \kappa C$ is the (voltage controlled) coupling capacitance, with $\kappa \ll 1$ being the coupling strength coefficient. Using the complex-mode representation for each resonator $n$, we rewrite the circuit equations Eq. (S5) as

$$i\begin{pmatrix}\dot{\psi}_1^*\\ \dot{\psi}_2^*\end{pmatrix} = \begin{pmatrix}\omega_0\left(1-\frac{\kappa}{2}\right) + i\frac{\Gamma_1^{(0)}}{2} - i\chi|\psi_1|^2 & \omega_0\kappa/2\\ \omega_0\kappa/2 & \omega_0\left(1-\frac{\kappa}{2}\right) - i\frac{\Gamma_2^{(0)}}{2} - i\chi|\psi_2|^2\end{pmatrix}\begin{pmatrix}\psi_1^*\\ \psi_2^*\end{pmatrix}; \quad (S6)$$

where we have invoked, as in the case of single RLC resonator, the rotating wave approximation, together with the weak coupling limit $\kappa \ll 1$ and high-Q $\frac{\Gamma_n^{(0)}}{\omega_0} \ll 1$ approximations. We further simplify Eq. (S6) by defining the rescaled complex field $a_n = \sqrt{\chi/\omega_0}\,\psi_n^*$ and time $\tau = \omega_0 t$,

$$i\frac{d}{d\tau}\begin{pmatrix}a_1\\ a_2\end{pmatrix} = \begin{pmatrix}\left(1-\frac{\kappa}{2}\right) + i\left(\gamma_1^{(0)} - |a_1|^2\right) & \kappa/2\\ \kappa/2 & \left(1-\frac{\kappa}{2}\right) - i\left(\gamma_2^{(0)} + |a_2|^2\right)\end{pmatrix}\begin{pmatrix}a_1\\ a_2\end{pmatrix}; \quad (S7)$$

where $\gamma_n^{(0)} = \Gamma_n^{(0)}/(2\omega_0) = 1/(2R_n^{(0)}C\omega_0)$.

**Coupling of the circuit to transmission lines**

The effect of a weak coupling of the RLC resonators to a transmission line (TL) via a coupling capacitance $C_e = \varepsilon C$, $\varepsilon \ll 1$, can be also modeled using CMT. At the node connecting the TL with the coupling capacitance, the voltage and current flowing toward the $n-$th ($n = 1,2$) RLC resonator can be written as a superposition of forward and backward propagating voltage waves, $V_{TL,n}^{(+)}$ and $V_{TL,n}^{(-)}$

$$V_{TL,n} = V_{TL,n}^{(+)} + V_{TL,n}^{(-)}; \quad I_{TL,n} = I_{TL,n}^{(+)} + I_{TL,n}^{(-)} = \frac{1}{z_0}\left(V_{TL,n}^{(+)} - V_{TL,n}^{(-)}\right), \quad (S8)$$

where $z_0 = 50\Omega$ is the TL's characteristic impedance. In turn, the voltages can be represented by complex wave amplitudes $V_{TL,n}^{(\pm)} = \sqrt{\frac{z_0}{2}}\left(S_n^{(\pm)} + S_n^{(\pm)*}\right)$, where $S_n^{(\pm)} = \left|S_n^{(\pm)}\right|e^{-i\omega_0 t}$, being $\left|S_n^{(\pm)}\right|$ a slowly varying amplitudes. For a single RLC resonator which is weakly coupled to a TL, we have

$$I_{TL,n} = C_e(\dot{V}_{TL,n} - \dot{V}_n); \quad -\frac{1}{C}\dot{I}_{TL,n} + \ddot{V}_n + \Gamma_n^{(0)}\dot{V}_n + \omega_0^2 V_n = 0, \quad (S9)$$

where $V_n$ represents the voltage at the $n$-th RLC resonator with characteristic frequency $\omega_0$ and decay rate $\Gamma_n^{(0)}$. Under the assumptions of impedance matching $z_0/Z_0 \sim O(1)$ and weak coupling $\varepsilon \to 0$, we rewrite Eqs. (S8) using the complex mode amplitude of the resonator $\psi^*$ and the input/output wave amplitude $S^{(\pm)}$

$$i\frac{d\psi_n^*}{dt} \approx \omega_0\left(1 - \sqrt{\frac{\eta\,Z_0}{2\,z_0}}\right)\psi_n^* - \eta\omega_0\psi_n^* - i\sqrt{2\omega_0\eta}S_n^{(+)}$$
$$S_n^{(-)} \approx S_n^{(+)} - i\sqrt{2\omega_0\eta}\psi_n^*, \quad (S10)$$

where we also invoked the rotating wave approximation, and we introduced the TL-RLC coupling coefficient $\eta \equiv (z_0/Z_0)\varepsilon^2/2$. Using the transformations $a_n = \sqrt{\chi/\omega_0}\,\psi_n^*$ and $\tau = \omega_0 t$, and combining Eqs. (S7) and (S10) we arrive to Eq. (1) of the main text which describes the whole system of coupled RLC resonators and TLs. Let us finally point out that when the input wave $S_n^{(+)} = 0$, the output power emitted from the $n$-th node takes a simple form

$$P_n = \left|S_n^{(-)}\right|^2 \approx 2\omega_0\eta|\psi_n^*|^2 = 2\chi\eta|a_n|^2, \quad (S11)$$

**Equations of motion in polar form**

It is convenient for our analysis, to rewrite Eq. (1) of the main text in polar representation. To this end, we express the complex amplitudes as $a_n(\tau) = A_n(\tau)e^{i\varphi_n(\tau)}e^{-if\tau/f_0}$ where the magnitudes $A_n \geq 0$ and the phases $\varphi_n$ of the fields in resonator $n = 1,2$ are real numbers. Substitution of these expressions back to Eq. (1) leads to the following set of coupled differential equations

$$\dot{A}_1 = \left(\gamma_1^{(0)} - A_1^2 - \eta\right)A_1 + \left(\frac{\kappa}{2}\right)A_2 \sin\varphi; \quad \dot{A}_2 = -\left(\gamma_2^{(0)} + A_2^2 + \eta\right)A_2 - \left(\frac{\kappa}{2}\right)A_1 \sin\varphi$$

$$A_1\dot{\varphi}_1 = \left(\frac{f}{f_0} - \nu_k\right)A_1 - \left(\frac{\kappa}{2}\right)A_2 \cos\varphi; \quad A_2\dot{\varphi}_2 = \left(\frac{f}{f_0} - \nu_k\right)A_2 - \left(\frac{\kappa}{2}\right)A_1 \cos\varphi$$

(S12)

where $\varphi \equiv \varphi_2 - \varphi_1$ is the relative phase. If $A_1, A_2 \neq 0$, the last two equations can be combined in a compact form

$$\dot{\varphi} = \frac{\kappa}{2}\left(\frac{A_2}{A_1} - \frac{A_1}{A_2}\right)\cos\varphi, \quad (S13)$$

Finally, in this polar representation the emitted power spectrum Eq. (S10) takes the form $P_n(\omega) = \eta \cdot A_n^2 \left(\frac{4}{3}\frac{1}{bZ_0^2}\right)$.

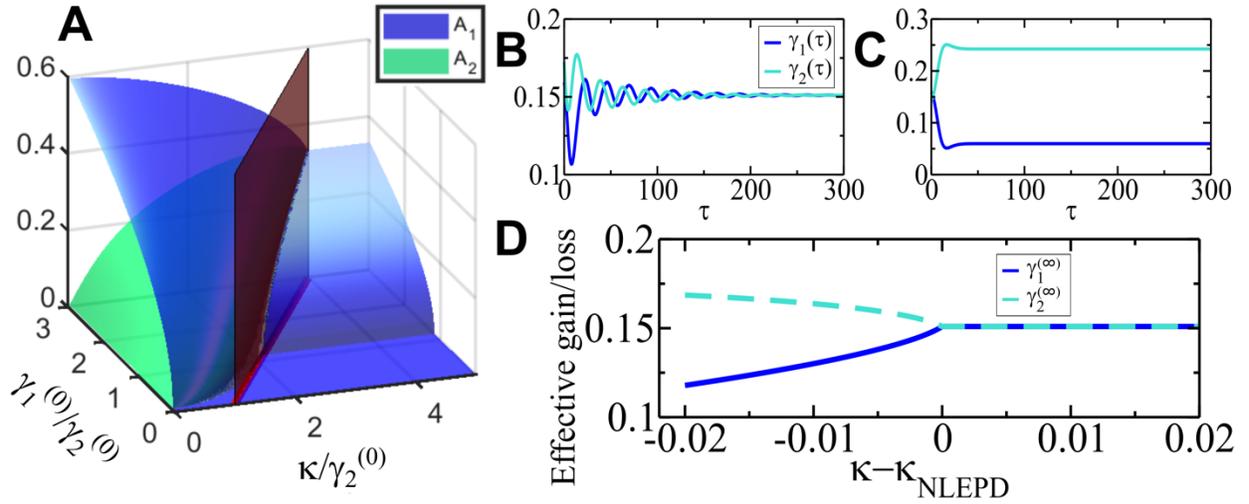

Figure S1: Nonlinear Super-modes and effective gain/loss coefficient: (A) The structure of the NS in the $(\kappa/\gamma_2^{(0)}, \gamma_1^{(0)}/\gamma_2^{(0)})$ parameter space. Blue (green) surface indicates the value of the field amplitude in the first $A_1$ (second $A_2$) RLC resonator. Temporal evolution of the effective gain (loss) coefficients $\gamma_1(\tau)$ ($\gamma_2(\tau)$) for linear gain/loss parameter values $\gamma_1^{(0)} \approx 0.18, \gamma_2^{(0)} \approx 0.12$ and (B) $\kappa = \kappa_{NLEPD} + 0.1$ in the AD regime where the system is protected by a parity-time symmetry; (C) $\kappa = \kappa_{NLEPD} - 0.1$ in the OD regime where the system is explicitly violates the parity-time symmetry. (D) Overview of the asymptotic gain/loss coefficients versus the coupling variation from the NLEPD. The coupling to the TL is taken $\eta = 0.01$ while $\kappa_{NLEPD} \approx 0.3$.

**Nonlinear supermodes**

The nonlinear supermodes correspond to the fixed points of Eqs. (S12,S13). We find the fixed points by requiring constant magnitudes and relative phase, i.e., $\dot{A}_1 = \dot{A}_2 = 0$, and $\dot{\varphi} = 0$. The latter condition leads to three possible scenarios: (1) $A_1 = A_2 \neq 0$, (2) $\cos\varphi = 0$, and (3) a variant of the first case where $A_1 = A_2 = 0$. We refer to this last case as the trivial scenario and it is not relevant for hypersensitive sensing schemes. Therefore, we will not analyze it further. A panorama of the nodal amplitudes $(A_1, A_2)$ is shown in Fig. S1A versus the relative gain $\gamma_1^{(0)}/\gamma_2^{(0)}$ and the relative capacitive coupling $\kappa/\gamma_2^{(0)}$. Below, we analyze in detail each one of the first two cases.

*AD Supermodes*

We consider first the case of identical field amplitude in both resonators, i.e., $A_1 = A_2 = A$. This corresponds to the AD domain, characterized by an exact parity-time symmetry. From the first set of Eqs. (S12), we find that the field amplitude and the relative phase are given by the expressions

$$A = \sqrt{\frac{\gamma_1^{(0)} - \gamma_2^{(0)} - 2\eta}{2}}, \qquad \sin\varphi = -\frac{\gamma_1^{(0)} + \gamma_2^{(0)}}{\kappa}, \qquad (S14)$$

which, in turn, establishes bounds for the parameters where such a solution exists. From the second of these equations, we conclude that the two supermodes differ in their relative phase i.e. $\varphi_\pm = \left(\frac{-1\pm 1}{2}\right)\pi \mp \sin^{-1}\left(\frac{\gamma_1^{(0)}+\gamma_2^{(0)}}{\kappa}\right)$ while the coupling between resonators is bounded by $\kappa \geq \gamma_1^{(0)} + \gamma_2^{(0)}$. This condition determines the boundary between the AD and the OD domains (regions II and I, respectively in Fig. 1B). On the other hand, the physical requirement that $A \in \mathcal{R}_{>0}$ which guarantees the existence of a non-trivial steady state, leads to the constraint that the gain must be strong enough to overcome the total loss of the system, i.e. $\gamma_1^{(0)} \geq \gamma_2^{(0)} + 2\eta$. This condition determines the boundary between the AD domain and region III (see Fig. 1 B).

The nonlinear eigenfrequencies $f_\pm$ associated with the NS of Eq. (S14) can be found by imposing in either of the second set of Eqs. (S12), the fixed-point condition $\dot{\varphi}_{1,2} = 0$ together with the expression for $\cos\varphi_\pm = \pm\sqrt{1 - \left(\frac{\gamma_1^{(0)}+\gamma_2^{(0)}}{\kappa}\right)^2}$ derived from the second equation in Eq. (S14). We have for the corresponding frequencies

$$f_\pm = f_0 \cdot (\nu_\kappa \pm \tfrac{1}{2}\sqrt{\kappa^2 - \left(\gamma_1^{(0)} + \gamma_2^{(0)}\right)^2}), \quad (S15)$$

which indicate the existence of a NLEPD at $\kappa_{NLEPD} = \gamma_1^{(0)} + \gamma_2^{(0)}$ corresponding to the transition between AD and OD domain, where not only the eigenfrequencies but also the eigenvectors coalesce.

*OD Supermodes*

Next, we discuss the fixed points of Eqs. (S12,S13) when $\cos\varphi = 0, i.e. \varphi = \pm\pi/2$. From the first set of Eqs. (S12), we conclude that only the relative phase $\varphi = -\frac{\pi}{2}$ ensures a solution with $A_2^2 \geq 0$. In this case the NS have an asymmetric field amplitude, i.e., $A_1 \neq A_2$ and the corresponding field intensities at each resonator are

$$A_1^2 = \gamma_1^{(0)} - \eta - \frac{\kappa}{2}\rho; \quad A_2^2 = -\gamma_2^{(0)} - \eta + \frac{\kappa}{2}\rho, \quad (S16)$$

where, $\rho = A_2/A_1 > 0$ represents the relative field amplitude and is a solution of the following quartic equation

$$0 = 1 - 2\rho\left(\frac{\gamma_2^{(0)}+\eta}{\kappa}\right) - 2\rho^3\left(\frac{\gamma_1^{(0)}-\eta}{\kappa}\right) + \rho^4. \quad (S17)$$

Out of the four roots of Eq. (S17), one has to select the ones that satisfy (a) $\rho \in \mathcal{R}_{>0}$, , (b) $\rho \leq 2\frac{(\gamma_1^{(0)}-\eta)}{\kappa}$ such that $A_1^2 \geq 0$, and (c) $\rho \leq \frac{\kappa}{2(\gamma_2^{(0)}+\eta)}$ such that $A_2^2 \geq 0$ (see Eq. (S16)). It turns out from our extensive numerical analysis that the requirement for the stability of the fixed point in the OD domain, results in $\rho \leq 1$ for the relative field amplitude of the nonlinear supermode Eq. (S16). This output is consistent with the intuition that the intensity at the lossy resonator is smaller than the intensity at the gain resonator at the OD domain where parity-time symmetry is explicitly violated.

The corresponding nonlinear eigenfrequency $f$ can be found by substituting in the last Eq. (S12) the value of the relative phase $\varphi = -\frac{\pi}{2}$. In this case, we get

$$f = f_0 \cdot \nu_\kappa. \quad (S18)$$

*Symmetry phases of the dimer system of Eq. (1)*

It is finally instructive to evaluate the effective gain and loss coefficients $\gamma_1(t)$ and $\gamma_2(t)$ as a function of time in the AD and OD regimes. In Fig. S1B we show their temporal behavior for a typical set of parameters $\gamma_1^{(0)} \approx 0.18$, $\gamma_2^{(0)} \approx 0.12$ and $\kappa = \kappa_{NLEPD} + 0.1$ for which the system of Eq. (1) is in the AD domain ($\kappa_{NLEPD} \approx 0.3$). We see that in the asymptotic time limit, $\gamma_1^{(\infty)} = \gamma_2^{(\infty)}$, indicating that the electronic dimer is in the exact parity-time symmetric phase. Similarly, in Fig. S1C we show the temporal behavior of $\gamma_1(t)$ and $\gamma_2(t)$ for the same values of parameters $\gamma_1^{(0)}$, $\gamma_2^{(0)}$ as previously and $\kappa = \kappa_{NLEPD} - 0.1$ for which the system of Eq. (1) is in the OD domain. In this case, the asymptotic values of the effective gain and loss parameters, differ from one-another, i.e., $\gamma_1^{(\infty)} \neq \gamma_2^{(\infty)}$, indicating that the system is in an explicitly broken parity-time symmetric phase. A panorama of the asymptotic values $\gamma_1^{(\infty)}, \gamma_2^{(\infty)}$ versus the coupling constant $\kappa$ for a fixed relative gain $\frac{\gamma_1^{(0)}}{\gamma_2^{(0)}} = 1.5$ is shown in Fig. S1D.

**Jacobian matrix and stability analysis in the parameter space**

The system of Eqs. (S12,S13) can be written in the form $\frac{d\vec{u}}{d\tau} = \vec{f}(\vec{u})$, where $\vec{u} = (A_1, A_2, \varphi)^T$. The NSs of the previous section $\vec{u}_0$ are obtained by the fixed-point condition $\frac{d\vec{u}_0}{d\tau} = \vec{f}(\vec{u}_0) = 0$. Linearizing the equations of motion around $\vec{u}_0$ we get $\vec{f}(\vec{u}_0 + \delta\vec{u}) \approx \hat{J}(\vec{u}_0)\,\delta\vec{u}$, where $\hat{J}(\vec{u}_0)$ is the Jacobian matrix evaluated at $\vec{u}_0$. Subsequently the linearized equations of motion read as,

$$\frac{d\delta\vec{u}}{d\tau} = \hat{J}(\vec{u}_0)\,\delta\vec{u}; \quad \hat{J}(\vec{u}) = \begin{pmatrix} \gamma_1^{(0)} - \eta - 3A_1^2 & \frac{\kappa}{2}\sin\varphi & \frac{\kappa}{2}A_2\cos\varphi \\ -\frac{\kappa}{2}\sin\varphi & \gamma_2^{(0)} - \eta - 3A_2^2 & -\frac{\kappa}{2}A_1\cos\varphi \\ -\frac{\kappa}{2}\cos\varphi\left(\frac{A_2}{A_1^2} + \frac{1}{A_2}\right) & \frac{\kappa}{2}\cos\varphi\left(\frac{1}{A_1} + \frac{A_1}{A_2^2}\right) & \frac{\kappa}{2}\sin\varphi\left(\frac{A_1}{A_2} - \frac{A_2}{A_1}\right) \end{pmatrix}. \quad (S19)$$

The first Lyapunov criterion allow us to characterize the stability of each nonlinear supermode by analyzing the eigenvalues of the Jacobian matrices evaluated at the NSs $\vec{u}_0$. The fixed-point solution $\vec{u}_0$ is stable if all the eigenvalues $\{\lambda_n; n = 1,2,3\}$ of the Jacobian (evaluated at the supermode) have negative real part, indicating that the solution is an attractor. If, on the other hand, one of the eigenvalues of the Jacobian has positive real part, the solution $\vec{u}_0$ is unstable. In both cases the associated fixed points are classified as hyperbolic equilibria. These equilibria points are further classified as nodes or foci depending on if the corresponding Jacobian eigenvalues are purely real or have also an imaginary component. In case that one of the eigenvalues of the Jacobian has a zero real part, then the fixed point is characterized as non-hyperbolic. Non-hyperbolic equilibria are not robust to small perturbations (i.e., the system is not structurally stable) and their stability is determined via direct dynamical simulations with the initial dynamical equations.

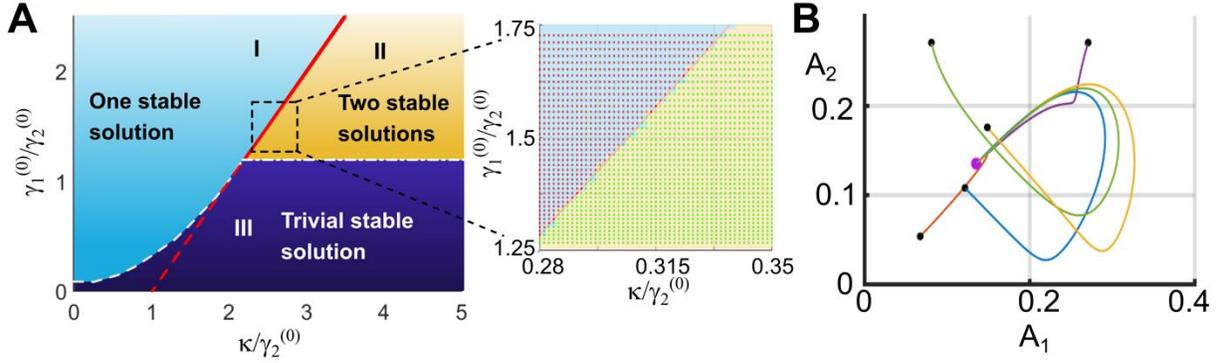

Figure S2: **Parameter space characterization and stability of fixed points.** (A) A characterization of the parameter space $(\kappa/\gamma_2^{(0)}, \gamma_1^{(0)}/\gamma_2^{(0)})$ of the nonlinear dimer described by the TCMT Eq. (S12,S13) in terms of the nature and number of its (stable) fixed points. Domain I is identified with the OD phase and supports one nontrivial stable fixed point; Domain II is identified with the AD phase and supports two non-trivial stable fixed points; Domain III supports only trivial stable fixed points (i.e., $A_1 = A_2 = 0$). The white dashed lines indicate the borders of various domains (see main text). The red dashed line indicates NLEPDs associated with the coalescence of unstable fixed points and occur in the domain III. The red solid line indicates NLEPD associated with the coalescence of stable fixed points and occur at the transition between OD and AD. The inset indicates the parameter domain for which we have analyze the stability (and number) of steady-state solutions of the actual electronic circuit using an NGSPICE simulator. The blue highlighted area indicates the OD domain while the yellow highlighted area indicates the AD domain. The red line between the two domains matches with the one found from the analysis of the TCMT. (B) Dynamical simulations with Eqs. (S12,S13) with $\kappa = \kappa_{NLEPD}$ indicate that various initial conditions (black circles) in the phase space of the system, converge asymptotically to the corresponding fixed point (violet circle).

A panorama of the $(\kappa, \frac{\gamma_1^{(0)}}{\gamma_2^{(0)}})$ parameter space and its partition to various domains, according to the nature (stable nontrivial versus stable trivial) and number (one, two or none) of stable fixed points, is shown in Fig. S2A. This map has been created by analyzing the stability of the TCMT Eq. (1) of the main text (Eqs. (S12,S13) of the Supplementary Information) together with the evaluation of the eigenmodes of the Jacobian matrix Eq. (S19). In the inset of Fig. S2A, we show the results of the numerical analysis using NGSPICE. We have simulated the evolution of several initial conditions for various $(\kappa, \frac{\gamma_1^{(0)}}{\gamma_2^{(0)}})$ values in the domain between AD and OD (dashed square) and analyzed the emitted power spectrum associated with the voltage of the gain resonator. The number of extracted frequencies peak(s) and the values of the amplitudes $A_1, A_2$ have been used to identify the parameter domain.

Let us discuss in more detail the stability of the NLEPD at the transition between AD and OD domains, see Fig. S2A. This fixed point turns out to be non-hyperbolic (i.e. one eigenvalue of the Jacobian matrix has zero real part). For an analysis of its stability, we cannot rely on the first Lyapunov criterion. Instead, we have performed direct dynamical simulations for a large ensemble of initial conditions in the phase space of the system and analyze their long-time dynamics. We have found that in all cases, the trajectories are attracted to a final state that is the NLEPD, see Fig. S2B.

For completeness of the discussion, we would also like to comment on the unstable non-trivial solutions that exists both in the AD domain and OD domain (see dashed lines in Fig. 2A). These are fixed points of Eq. (S12) corresponding to $\cos\varphi = 0$ and $(A_1, A_2) \neq (0,0)$ and exists both in the AD and in the OD domains. The analysis of these fixed points follows the lines of the "OD Supermode" section and correspond to roots of Eq. (S17) with $\rho > 1$. The corresponding frequency of these solutions is given by the either of the second set of Eq. (S12) and are $f_{unstable} = f_0 \cdot \nu_k$.